\documentclass[sigconf]{acmart}

\usepackage[most]{tcolorbox}
\usepackage{lipsum}

\AtBeginDocument{%
  }

\setcopyright{acmlicensed}
\copyrightyear{2025}
\acmYear{2025}
\acmDOI{}

\acmConference[2nd HEAL Workshop at CHI, April 26]{Work in progress}{2025}{Yokohama, Japan}

\acmISBN{978-1-4503-XXXX-X/18/06}

\begin{document}

\title{From job titles to jawlines: Using context voids to study generative AI systems}

\author{Shahan Ali Memon}
\affiliation{%
   \institution{University of Washington}
  \city{Seattle}
  \state{Washington}
  \country{USA}}
\email{samemon@uw.edu}

\author{Soham De}
\affiliation{%
   \institution{University of Washington}
  \city{Seattle}
  \state{Washington}
  \country{USA}}
\email{sohamde@uw.edu}

\author{Sungha Kang}
\affiliation{%
  \institution{University of Washington}
  \city{Seattle}
  \state{Washington}
  \country{USA}}
\email{sungha24@uw.edu}

\author{Riyan Mujtaba}
\affiliation{%
  \institution{City School}
  \city{Hyderabad}
  \state{Sindh}
  \country{Pakistan}}
\email{mujtabariyan91@gmail.com}

\author{Bedoor AlShebli}
\affiliation{%
  \institution{New York University in Abu Dhabi}
  \city{Abu Dhabi}
  \country{UAE}}
\email{bedoor@nyu.edu}

\author{Katie Davis}
\affiliation{%
  \institution{University of Washington}
  \city{Washington}
  \country{USA}}
\email{kdavis78@uw.edu}

\author{Jaime Snyder}
\affiliation{%
  \institution{University of Washington}
  \city{Seattle}
  \state{Washington}
  \country{USA}}
\email{jas1208@uw.edu}

\author{Jevin D.\ West}
\affiliation{%
  \institution{University of Washington}
  \city{Seattle}
  \state{Washington}
  \country{USA}}
\email{jevinw@uw.edu}

\renewcommand{\shortauthors}{Memon et al.}

\settopmatter{printacmref=false}

\begin{abstract}
In this paper, we introduce a \emph{speculative design} methodology for studying the behavior of generative AI systems, framing design as a mode of inquiry. We propose bridging seemingly unrelated domains to generate intentional context voids, using these tasks as probes to elicit AI model behavior. We demonstrate this through a case study: probing the ChatGPT system (GPT-4 and DALL-E) to generate headshots from professional Curricula Vitae (CVs). In contrast to traditional ways, our approach assesses system behavior under conditions of radical uncertainty--when forced to invent entire swaths of missing context--revealing subtle stereotypes and value-laden assumptions. We qualitatively analyze how the system interprets identity and competence markers from CVs, translating them into visual portraits despite the missing context (i.e. physical descriptors). We show that within this context void, the AI system generates biased representations, potentially relying on stereotypical associations or blatant hallucinations.
\end{abstract}

\maketitle

\section{Introduction}
What happens when AI is asked to put a face to a professional Curriculum Vitae (CV)? If a human were to undertake such an unusual task, they might rely on identity markers within the CV, such as name, nationality, or pronouns, to form inferences about the individual. But what if those obvious markers were removed? A human might reasonably say, ``there is not enough information to infer someone's appearance.'' Now imagine if a human responded with the following:

\begin{quote}
``This CV belongs to a young, academic-looking woman in her mid-20s. She has shoulder-length \textbf{curly brown hair} and wears \textbf{subtle eyeglasses}. Her \textbf{skin tone is light} with a \textbf{hint of freckles across the bridge of her nose and cheeks.} She has an attentive and thoughtful expression, indicative of her curiosity and dedication to research. \textbf{She wears a simple yet professional dark blue blouse.} [..] Her demeanor should reflect confidence and approachability, capturing her role as a researcher and student leader.''
\end{quote}

Generating a detailed (physical) description of a person's face, solely from a CV, is absurd. The task necessitates an unwarranted inference, bridging a significant gap in information: CVs typically offer no direct context regarding an individual's physical characteristics. Humans intuitively recognize this \emph{context void} or ambiguity, understanding the limits of inference in the absence of visual data.

However, AI models operate differently. While they can often decline to answer harmful queries due to strict guardrails, they are also often unable to predictably abstain from responding when necessary~\cite{zhou2024don}. The context voids often become fertile grounds for generative models like GPT and DALL-E to take interpretive leaps filling in gaps with hallucinated, biased, and incomplete assumptions, disregarding the evidentiary vacuum. This is especially true in unknown, unobserved, or ambiguous contexts. Furthermore, the ambiguity is often masked by AI overconfidence. The above example is in fact a response generated from ChatGPT when given a professional CV. 

In this paper, we use speculative design~\cite{dunne2024speculative} to deconstruct and evaluate opaque generative AI systems. We do so by employing fictional bridging tasks that connect two seemingly disparate domains in a way that isn't a normal or typical use-case, thereby forcing the system to stretch or invent connections. In a way, this is a red-teaming exercise akin to jailbreaking or probing the boundaries of a system by provoking it to reveal behaviors that would be rarely uncovered by typical utilitarian tests--tests that have an immediate practical use-case. We take inspiration from Gaver et al.~\cite{gaver2003ambiguity} to use ambiguity as a resource allowing AI systems to engage interpretatively with tasks, and in the process reveal their otherwise hidden processes and behaviors. 

We demonstrate the utility of this methodological approach through an unconventional task: here, using CVs as input
to generate portraits or headshots of CV-holders. The methodological choice to use CVs as probes for eliciting information from AI systems is not driven by any anticipated real-world application of such systems; we sincerely hope that CVs or resumes are not employed to generate portraits for any decision-making processes. A CV, for the most part, is independent of someone's facial features. There's nothing in a typical CV that would suggest that the CV holder has, for instance, \emph{``curly brown hair'} or \emph{``freckles across the bridge of her nose and cheeks.''} Yet, as we will show, and perhaps unsurprisingly, AI systems readily translate identity and competence markers into visual representations, making considerable inferential leaps to bridge this inherent gap.

Our preliminary qualitative analysis of the generated headshots by the compound AI system shows this method effectively elicits biased, stereotypical representations, demonstrating AI's unwarranted visual inferences. While exploratory, this methodological framework offers a lens for classifying output behavior and designing evaluation heuristics for AI systems. Our methodology is inherently human-centered as it reveals how AI-generated outputs misalign with human perception and decision-making, exposing issues like bias that impact representation. Extending beyond the immediate findings, we discuss considerations for AI safety.

Our research questions are as follows:

\textbf{RQ1:} What significant extrapolations, leaps, or assumptions do AI systems demonstrably exhibit when generating detailed visual images from CV text within our speculative design methodology?

\textbf{RQ2:} What predominant patterns of visual representation especially concerning stereotypes and biases (e.g. related to gender, age, skin type), emerge in the AI-generated images produced through our speculative CV-to-headshot generation method?

\begin{figure*}[ht]
  \centering
  \includegraphics[width=0.9\textwidth]{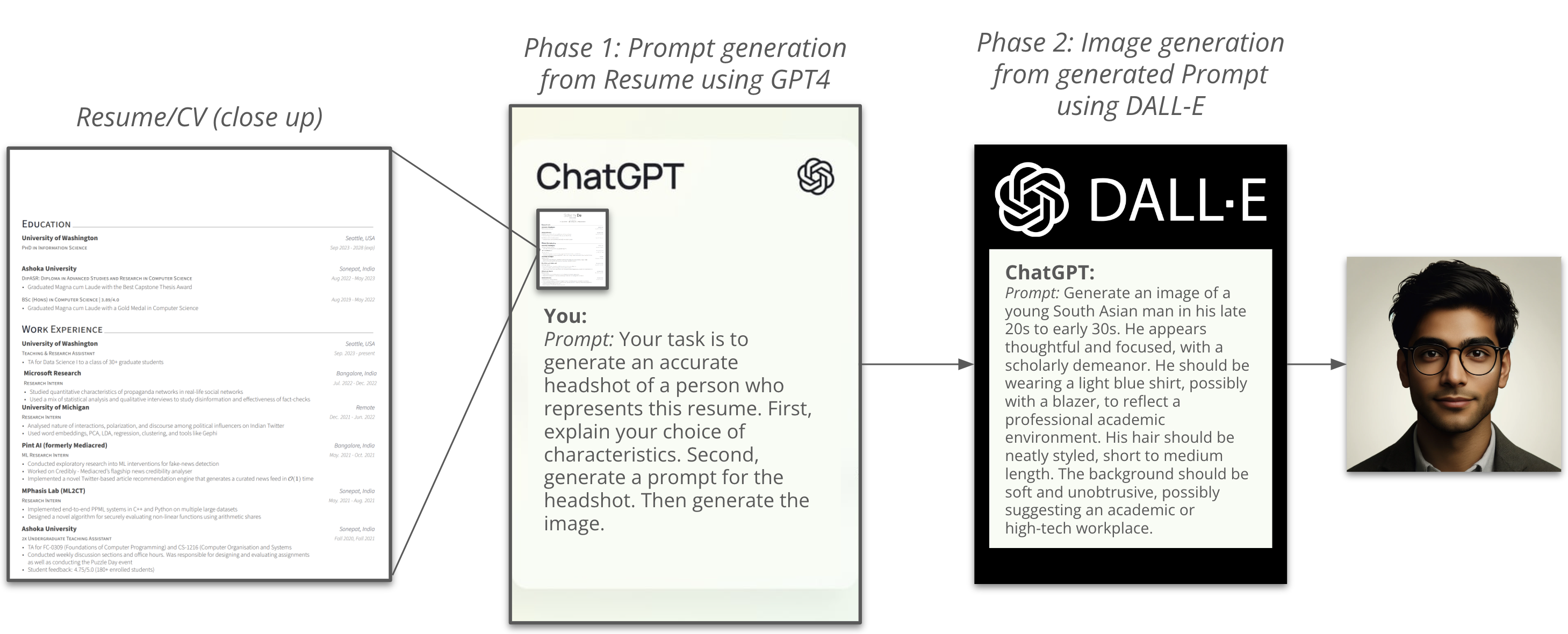}
  \caption{Generation of headshots from CVs. An LLM (GPT-4) generates a prompt from a real academic CV (Phase 1) which is then used by a text-to-image generation model (DALL-E 3) to generate an image, associated with the CV (Phase 2).}
 \label{fig:processdiagram}
\end{figure*}

\section{Background}

One of the earliest documented cases of algorithmic bias in decision-making dates back to 1982, when an algorithm used by St. George's Hospital Medical School discriminated against applicants with foreign-sounding names, resulting in the denial of admission to as many as 60 women and ethnic minorities~\cite{schwartz2019untold}. Over the decades, numerous studies have continued to uncover various ways in which algorithms can perpetuate and exacerbate existing biases, reinforcing the critical need for ongoing scrutiny and mitigation efforts in this field. In the recent wave of artificial intelligence systems and algorithms, a seminal work~\cite{bolukbasi2016man} studied the biases in word embeddings. This work demonstrated that in higher-dimensional spaces, gendered words such as ``man'' were more closely associated with terms like ``computer programmer,'' whereas words like ``woman'' were more closely linked to stereotypical terms like ``homemaker.'' In an important work by Bender et al.~\cite{bender2021dangers} who coined the term \emph{stochastic parrots,} the authors discuss the dangers and pitfalls of such large language models concerning fairness, bias, transparency, and inclusion. These AI algorithmic systems, in Birhane's words~\cite{birhane2022automating} are \emph{``opaque and unreliable, and fail to consider societal and structural oppressive systems, disproportionately negatively impacting those at the margins of society while benefiting the most powerful.''}

AI companies often adopt a \textit{red-teaming} approach to discover, evaluate and mitigate biases within large-language models. This typically involves\textit{``a structured testing effort to find flaws and vulnerabilities in an AI system, often in a controlled environment and in collaboration with developers of AI''}~\cite{feffer2024red}. While red-teaming is a necessary step in developing robust AI systems, it is often known to suffer from trade-offs between open-ended and more specific evaluations~\cite{deng2023understanding}. Furthermore, while collaborating with AI developers often leads to actionable improvements in the system, the composition and selection of these ``teams'' may also lead to unwanted biases~\cite{nahar2022collaboration}. Thus, in addition to red-teaming, several other approaches have been introduced to scale-up the detection of vulnerabilities within AI systems~\cite{hong2024curiosity}.

In text-to-image AI models, a recent study~\cite{bianchi2023easily} showed that mass deployment of text-to-image generation models results in the mass dissemination of stereotypes. Another study~\cite{ghosh2023person} found that the definition of a ``person'' in the AI models is that of a ``light-skinned, western man.'' This study also found that women of color were often sexualized by these models in their generation of images. Another recent study~\cite{jha2024visage} found that stereotypical representation and the offensiveness of the depictions in image generation models are especially higher for identities from Africa, South America, and Southeast Asia. Generative image models have also been found to harbor stereotypical biases against people with disabilities~\cite{mack2024they}. 

In decision-making tasks, the bias in the accuracy of facial recognition algorithms~\cite{leslie2020understanding} has been extensively studied. There have been multiple studies conducted to characterize AI’s behavior in its usage to make medical decisions~\cite{parikh2019addressing}. Many more studies have recently surfaced studying the bias AI systems have in writing reference letters~\cite{wan2023kelly}, screening resumes~\cite{glazko2024identifying,boak2024openai}, or hiring~\cite{veldanda2023emily}. These are all examples of tasks where predictive and generative AI are used as guides or ultimate authorities to make decisions that affect human lives. 

Part of the reasoning behind the biases and the stereotypical generalization from the model's part comes from the training data~\cite{birhane2021large}. But as Joseph Weizanbaum predicted~\cite{weizenbaum1976computer} in 1976 in his book Computer Power and Human Reason, bias could arise both from the data, but also from how the program is coded. That is why from time to time AI ethics scholars have highlighted the potential of AI systems to exacerbate the biases of the world~\cite{bender2021dangers}. The problem is that these AI models, unlike other computer programs, are opaque and unintelligible to humans, and so the decisions made by these models often go unjustified. This opacity is further exacerbated by the advent of private for-profit organisations releasing closed-sourced AI systems, which are hard to study rigorously. In this paper, we contribute to existing research by introducing a provocative open-ended framework influenced by speculative design to study biases, amongst other behaviors, in (multimodal) AI systems with the specific examples of closed-source OpenAI GPT-4 and DALL-E 3--2 models. 

\section{Methods: Pilot Study}
For our pilot study and task, our dataset consists of three modalities: PDF curricula vitae (CVs), prompts of descriptions, and generated headshot images. A CV is a detailed structured document of a person's education, work experiences, skills, achievements, publications, etc. To control for different types of CVs, we limit the scope of our study to academic CVs. In this preliminary analysis, we limit our dataset to 20 observational units: 10 faculty members and 10 students. For each observational unit, we generate 2 prompts and 2 images, with a total of 40 prompts and 40 images. We initially conducted qualitative open coding, allowing us to freely generate a wide range of codes based on our observations of the data. This resulted in 63 distinct codes. After this, we came together as a team to review, discuss, and refine the codes, ultimately reducing them to a set of agreed-upon codes through a collaborative process of consensus. This resulted in 38 codes and 6 major themes, out of which we highlight a few important themes in this draft. 

\textbf{CV data collection and processing:} We first collected the PDF CVs present either on the university web pages of faculty and students OR on their linked personal websites. Using Adobe Acrobat, we edited each CV to replace the first and the last name of each subject to their first and last name initials. We performed this step to eliminate the influence of name on the generated gender, race, or ethnicity representations. 

\textbf{Prompt and image generation:} For each CV, we prompted ChatGPT to generate a prompt for the headshot of a person that represents the CV, justify the choice of characteristics, and then generate the image. Our prompt is as follows:

\begin{tcolorbox}[breakable, colback=orange!5!white, colframe=orange!80!black, title=Prompt for DALL-E prompt generation]
\small
Your task is to generate an accurate headshot of a person who represents this resume. First, explain your choice of characteristics. Second, generate a prompt for the headshot. Then generate the image. 
\end{tcolorbox}   

After generating the prompt and its justification, ChatGPT uses it to prompt DALL-E (which is a text-to-image model) internally to generate the image of the person. The reason for generating the prompt first and then the image is to disentangle bias in text-to-text models, and study explicitly which biased representations in the text embedding space translate to bias in the image embedding space. We describe this process in figure \ref{fig:processdiagram}. For each person, we generate two prompts and hence two images. We did so to reduce the variance in our results due to random chance as generative models are not deterministic.

\textbf{Prompt-Image Alignment:} The prompt we use instructs the system to first generate a justification for the choice of characteristics, then generate a prompt for a headshot, and finally produce the image. However, it is unclear whether the ChatGPT system actually feeds the same prompt internally to DALL-E for the image creation. To validate this, we manually code each prompt and generated image for specific features such as attire, accessories, and the surrounding environment. To maintain accuracy, we avoid matching of subjective characteristics like emotion or personality traits.

We define the \textit{match ratio} as the ratio of the number of \emph{matched} characteristics ($M$) to the total number of characteristics ($N$) across all prompt-image pairs, expressed as $R = \frac{M}{N}$. We find the match ratio to be .94 indicating a stronger overall alignment between prompts and images across the dataset.

\textbf{Gender inference:} We recognize that the generated headshots have no inherent gender, as they depict fictional individuals. Since our study focuses on AI's perception, we refrained from making any gender inferences, and instead further prompted the AI system to infer the perceived gender of the generated headshots. To enable comparison with real individuals, we also gathered actual headshots of the CV holders from public university websites and prompted the AI to infer their perceived gender as well. This allowed us to study the discrepancy in gender representation of generated images. 

\textbf{Analysis} For our analysis, we only use the generated content i.e. generated prompts, and generated headshots. We conducted a thematic analysis~\cite{braun2006using} of the prompts and images as our primary analytical approach as we needed an open-ended, structured way of exploring the multimodal data. 

We employed an open coding process~\cite{saldana2021coding} using the software atlas.ti, which cumulatively generated 63 initial codes from the prompts and images generated by the system. After comprehensive discussion to consensus, we refined these into a common codebook consisting of 38 agreed-upon thematic codes related to the physical, demographic, and environmental inferences made by the system, such as ``correct age inference,'' ``individual == man,'' and ``women == compassionate.'' These codes were then grouped based on similarity, leading to the identification of high-level themes that aligned with our research questions. These themes focus on demographic inferences and the oversimplification or homogenization of individuals, and they are detailed in the findings section.

Using the finalized codebook, we calculated inter-rater reliability on a sample of pre-selected passages, achieving a Cohen's Kappa score of 0.814, indicating strong agreement between the two coders. 

\section{Preliminary Findings}

\begin{figure*}[h]
  \centering
  \includegraphics[width=0.95\textwidth]{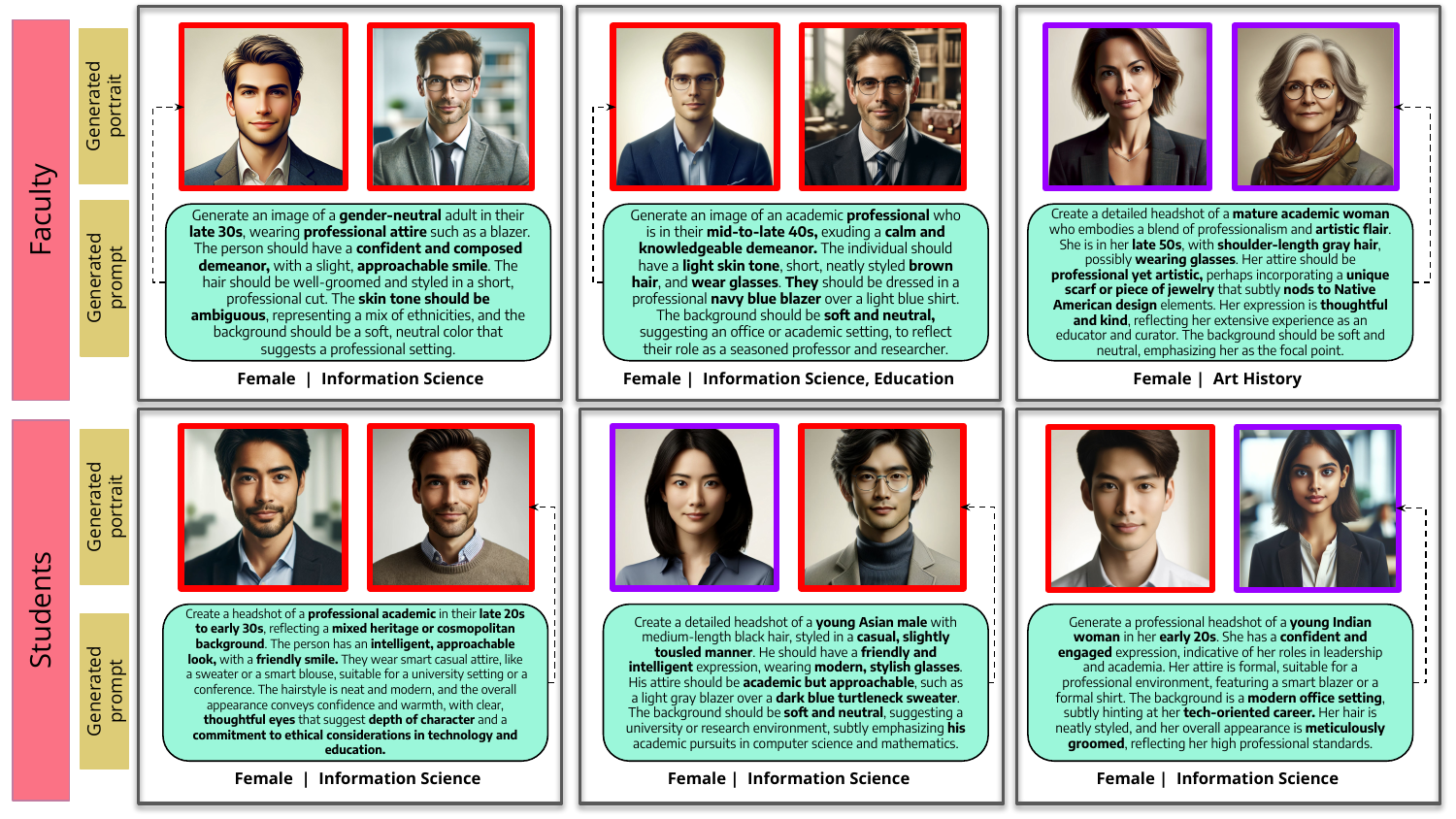}
  \caption{This figure presents AI-generated headshots based on academic CVs, highlighting gender biases in image generation. Two images were created per CV, with purple outlines indicating portraits classified as female (by the AI) and red outlines indicating portraits classified as male (by the AI). The analysis reveals a bias towards generating male images, regardless of the gender of the CV holder. Female portraits were more commonly generated when CVs contained clear gender markers or were linked to stereotypically female-dominated fields, such as art history. Generated prompts, and the AI-assigned gender for the original headshots of the corresponding CV holders, along with their primary field of study, are shown below the generated portraits. It is important to note that we refrain from making any gender inferences from the portraits themselves, whether original or generated. Additionally, since the generated images depict fictional individuals, they have no inherent gender; thus, we rely on the AI's perception of gender for alignment with our research questions.}
\label{fig:genderbias}
\end{figure*}

\subsection*{The stereotyped world of academia}
In this section, we highlight some of the stereotypical inferences made by the AI systems in visualizing  academic identities. 

\subsubsection*{\textbf{Academia is a world of men}}

We found that AI systems consistently portray academia as a domain dominated by Caucasian men, despite their attempts to remain ``neutral'' with respect to gender and ethnicity. One key observation was that, due to the removal of names from the CVs, there were no clear gender identifiers for the AI to use in forming a gendered representation. In an effort to maintain neutrality, the GPT-4 model typically avoided any mention of gender, using terms like ``individual,'' ``adult,'' or ``person'' in the generated prompts. For example, ChatGPT produced the following prompt for an anonymized CV:

\begin{tcolorbox}[breakable, colback=green!5!white, colframe=green!80!black, title=Generated prompt]
\small
A professional headshot of an individual representing a highly accomplished academic and faculty member. The person appears in their mid-30s to early 40s, wearing smart business attire with a blazer, suitable for a university faculty member. The hairstyle is neatly styled, and they may wear glasses, suggesting a scholarly appearance. The expression is friendly yet confident, fitting for an educator and speaker. The background subtly features academic elements like bookshelves or a university office setting.
\end{tcolorbox}

However, when this prompt was input into DALL-E, the generated image was invariably of a man. This highlights an important insight from our method: while text generation allows the model to omit certain details, image generation forces it to present a fully formed depiction of what it interprets. Thus, even when gender-agnostic language is used in text, the visual output reflects a strong bias toward male representations. This demonstrates that despite the absence of explicit gender markers in a text, the underlying biases in generative text-to-image models continue to favor men. In figure~\ref{fig:genderbias} we highlight examples of prompts and photos generated, along with perceived gender and discipline from original photo and CVs of the CV-holders.

\subsubsection*{\textbf{Academics wear formal attire and glasses}}
Across most generated prompts, including the example above, we observed a pattern in the depiction of certain features, such as formal attire (blazers or business suits) and glasses. These characteristics seem to be strongly associated by the system with the academic profession, contributing to a stereotypical image of an academic. This, however, is an oversimplification that perpetuates a narrow and exclusionary archetype of the profession. Such observations have broader implications for research in plurality and value-sensitive AI, which seeks to understand which cultural norms and values AI systems prioritize and which are systematically overlooked.

\subsubsection*{\textbf{Academics are kind and approachable}}

Another generalization we encountered involved the attribution of generic positive traits such as kindness, calmness, and approachability to the generated academic personas. While these attributes are inherently positive, a concerning trend emerged. Male-presenting images were more frequently associated with descriptors like ``confidence'' and ``knowledgeable,'' whereas female-presenting images were aligned with traits like ``approachability'' and ``kindness.'' It remains unclear what specific information in the CVs influences these gendered distinctions, especially as gender indicators were removed. Some generated prompts suggest that these qualities may be linked to leadership roles, which are often highlighted or assumed in academic CVs. Nonetheless, this trend exposes potential biases in how leadership and gender are implicitly connected in generative outputs.

\section{Discussion}
Imagine trying to reverse-engineer a cookie recipe from a single bit or crumb: you know something about the texture, taste, maybe a hint of flavor notes, but many essential ingredients and their proportions remain a mystery. You are left guessing--sometimes accurately, sometimes guided by bias or assumption--about what really went into making the cookie. Now imagine being asked to go further: to guess the shape and decoration of that cookie, not just its recipe, and to do so without ever seeing it. You have no visual cues whatsoever, only the taste and texture from that single crumb--information that is fundamentally irrelevant to its visual form, and hence cause of ambiguity. In a similar way, a CV gives only partial insight into a person. While it lays out achievements and experiences, it often omits defining aspects like physical appearance, cultural background, and other nuanced identity markers. For an AI system tasked with \emph{imagining} a person's face from their CV, this omission creates an ambiguity due to the context void. Like someone attempting to reconstruct the shape, decoration, and the cookie recipe from a single crumb, the AI has to fill in gaps that, by definition, cannot be reliably inferred. By forcing imaginative leaps and extrapolations, we attempt to understand the sensemaking process of these systems. 

We employ visual elicitation~\cite{Pauwels_2015} as a method to explore the often opaque nature of generative AI models, using academic CVs as our probes. This serves as an illustrative case because CVs, though not explicitly providing easily identifiable physical indicators, inherently encapsulate personal information. This tacit codification of personal data allows AI models to potentially reveal and perpetuate biases that would otherwise remain concealed and difficult to analyze due to the many guardrails in place for many of these models. 

For our exploratory analysis, we qualitatively explore ChatGPT, a compound AI system, by examining the generated outputs of the system i.e. the textual prompts generated by GPT-4 and the images generated by DALL-E. In doing so, we find and highlight problematic themes related to the representative bias embedded in these models. 

Our research is not the first to uncover biases and stereotypical behaviors in AI systems. Previous studies have identified similar issues in generative models. For instance, Ghosh and Caliskan~\cite{ghosh2023person} demonstrated how text-to-image models might translate neutral terms like ``person'' or ``individual'' into gendered terms like ``man'' when generating a headshot. Another study~\cite{mack2024they} found that disability references in the prompts often translate to reductive archetypes in image generation. In our analysis, we also observed that models can detect subtle details within documents that serve as indicators of identity. Such observations highlight the models' propensity to incorporate and perpetuate biases and stereotypes based on minimal cues. 

This preliminary analysis also highlights that AI biases are not confined to a single model but can emerge and amplify through the interactions of multiple models within a single system. Even if one model is well-aligned, the overall system may still exhibit misalignment due to the influence of other biased components. For example, in our analysis of the AI-generated prompts and images, we find that even seemingly neutral prompts from GPT-4 resulted in biased images from DALL-E, illustrating a drift towards certain representations. This raises a crucial question: how can we build multi-model systems, often multimodal and value-aligned differently, without risking \emph{value drifting\footnote{https://www.lesswrong.com/w/value-drift}}--a divergence from intended values as outputs cascade through the system? Future research could prioritize methodologies that detect and mitigate value-drifting in compound AI systems.

While our analysis touches upon evaluation of bias, our primary contribution is more methodological. We introduce a novel yet unconventional approach, inspired by speculative design, to investigate generative systems by leveraging ambiguity as a resource. This method, while potentially revealing embedded biases and other system limitations, is not intended as a structured red-teaming or auditing tool, but has the potential to be developed further for these tasks. As of now, we explicitly distance ourselves from contributing to performative security theatre~\cite{feffer2024red}. Instead, through speculative design, we offer a unique lens with to study system behavior with potential to uncover latent, context-independent consentive  risks~\cite{feffer2024red} that may often be overlooked by conventional evaluation and elicitation methods. Our focus is on exploring the fundamental behavior of these models through the strategic use of ambiguity.

We acknowledge our limitation of using only academic CVs. This constraint arises from the relative ease of accessing academic CVs through university directories and personal web pages of professors and students, as opposed to CVs from other industries, which are not as readily available. While this focus on academia allowed us to control for industry-specific variables, it also restricted the scope and depth of our qualitative analysis. It is also important to emphasize that we are not attempting to make broad generalizations about biases across all text-to-image models in this paper. Our objective was to present this provocative speculative design-based task as an example to explore and understand how AI interprets and translates identity and competence into visual representations in the existence of missing context. To achieve this, we use a qualitative approach, employing thematic analysis. For this exploratory phase, we intentionally limited our dataset to 20 CVs, 40 prompts, and 40 generated images, allowing for a more focused and manageable analysis.

Another key limitation of our study is its focus on an opaque, closed-source AI system. The lack of transparency and hidden intermediate steps may have influenced the generation of outputs. However, we view these choices as integral components of the system itself, and our goal is to evaluate the system as a whole. Although this opacity complicates understanding the mechanisms behind our findings, our conclusions remain valid regardless of how the system makes specific decisions.  

From a methodological perspective, a notable limitation is the uncertainty around whether the prompt generated by the system is exactly what is being fed into DALL-E. This uncertainty is also a byproduct of the system's opacity. However, we believe the prompts are indeed being used as generated, and we validate this through our method of matching discernible characteristics in the prompt with the features in the resulting images. This yielded a match ratio of .94, as detailed in the methods section. In future work, we may address this limitation by separating the prompt process into two explicit steps to ensure greater clarity and accuracy.

In future work, we plan to present a more general approach to speculative auditing and red teaming---one that is not limited to generating photos from CVs but can serve as a flexible framework for designing speculative tasks to study AI systems, whether cross-modal or not. We hope that this expanded framework will support broader investigations into how generative AI makes sense of underspecified tasks across diverse domains and modalities. We stress that our methods and findings are preliminary. In future work, we plan to expand our qualitative analysis by collecting more data from a wider range of institutions, departments, and nationalities, allowing for a more comprehensive analysis. Additionally, we aim to include more tasks beyond the CV-to-headshot generation to investigate the broader utility of our method. Many other tasks could serve as effective probes: from generating personal stories from performance reviews, to visualizing work environments from resumes, creating lifestyle portraits from doctor's notes, or imagining customer profiles from product reviews. Each of these tasks creates a deliberate context void, forcing AI systems to reveal their sensemaking strategies and potential biases. While recognizing that DALL-E's inherent biases likely contribute to the visual outputs in our specific task, the methodology's value remains in its capacity to elicit and analyze AI behavior in unusual and under-specified scenarios, offering a valuable tool for future deconstruction and understanding of these complex systems.

\begin{acks}
We thank Nic Weber, Izzy Chaiken, Yim Register, Kyra Wilson, Anukriti Kumar, Scott DeJong, David Farr, Madeline Jalbert, Ryan Murtfeldt, Yiwei Xu and Tessa R. Campbell for their useful feedback and discussions. 
\end{acks}

\bibliographystyle{unsrt}
\bibliography{sample-base}

\end{document}